\documentclass[fleqn,12pt]{revtex4}
\usepackage{graphicx}
\begin{document}
\title{Nuclear halo structure from quasielastic charge-exchange reactions}
\author{H.F. Arellano}
\email{arellano@dfi.uchile.cl}
\affiliation{
 Department of Physics, FCFM, University of Chile  \\
    Av. Blanco Encalada 2008, Santiago, Chile.}
\author{ W. G. Love}
\email{wglove@physast.uga.edu}
\affiliation{Department of Physics and Astronomy, 
     University of Georgia\\ Athens, GA-30605, USA.}

\begin{abstract}
Neutron and proton densities in the nuclear periphery are
investigated within (p,n) charge-exchange isobar transitions.
For this purpose we have developed parameter-free optical potentials~$^{1)}$
with a detailed treatment of the {\em in-medium} $t_{\tau}$ part of the
effective interaction.
Non local coupled-channel Lane equations are solved to obtain the
scattering observables.
The use of conventional proton and neutron densities 
significantly underestimates
Fermi (forward-angle) cross-sections 
in agreement with findings by various other groups.
However, we have found model-independent densities which provide
a remarkable improvement in the description of the quasielastic scattering data.
The densities obtained are consistent with recent measurements at CERN in
studies of the neutron-to-proton halo factor f(r)=Z$\rho_n/N\rho_p$ with
antiprotons~$^{2)}$.
These findings provide an alternative way to investigate the nuclear
periphery, and may also help to solve the long-standing puzzle of the
underestimated Fermi cross section in (p,n) charge-exchange phenomena.
\end{abstract}

\maketitle
%\begin{keyword}
%Neutron halo                         \sep 
%in-medium full-folding optical model \sep 
%quasielastic charge-exchange         \sep
%Lane equations
%\PACS 
%24.10.Eq \sep  %  Coupled-channel and distorted-wave models
%25.40.Kv \sep  %  Charge-exchange reactions
%21.30.Fe \sep  %  Forces in hadronic systems and effective interactions
%21.10.Gv       %  Mass and neutron distributions
%\end{keyword}

\section{INTRODUCTION}
An issue of prominent interest in recent nuclear research has been the
formation of neutron halo structures in various neutron-rich nuclei. Due to
its nature, this phenomenon is a clear manifestation of the isospin degrees
of freedom of the interacting nucleons in the nuclear medium. Since the
nucleon-nucleon interaction discriminates among the different isospin states
of the two nucleon system, its use within an \emph{in-medium} microscopic
approach for nucleon-nucleus collisions serves as a means to explore the
proton and neutron densities of the target ground state. In particular,
(p,n) charge-exchange reactions provide a rich and interesting arena for
exploring neutron-to-proton density differences.

The study of (p,n) charge-exchange reactions in the intermediate energy
range has been a subject of considerable attention during the past two
decades. These reactions are of great value in understanding the isovector
modes of excitations of the nucleus. At beam energies above 100 MeV, nucleon
charge-exchange reactions can be considered as a one step process thus
allowing a rather clean separation of the nuclear structure from the
underlying nucleon-nucleon effective interaction. In spite of these
advantageous considerations, no microcopic effort has been able to
satisfactorily describe the differential cross-section data without
phenomenological adjustments \emph{a posteriori}\cite{Bau01}.

\section{FRAMEWORK}
The optical potential for nucleon scattering and charge-exchange reactions
may be written in momentum space as: 
\begin{eqnarray}
\langle \mathbf{k^{\prime}} \nu^{\prime}\mu^{\prime}\mid U\mid \mathbf{k}
\nu \mu \rangle &=& \sum_{m,m^{\prime},n,n^{\prime}} \int\int d\mathbf{%
p^{\prime}} d\mathbf{p} \langle F \mid \psi_{ \textstyle{\frac{1}{2}}
m^{\prime},\frac{1}{2}n^{\prime}}^{\dagger}(\mathbf{p^{\prime}}) \overline{%
\psi}_{\frac{1}{2}m,\frac{1}{2}n}(\mathbf{p})\mid I\rangle \;  \nonumber \\
&& \langle \mathbf{p}\,^{\prime }m^{\prime }\,n^{\prime },\mathbf{k^{\prime}}%
\,\nu ^{\prime }\,\mu ^{\prime } \mid t\,\mid \mathbf{p}\,m\,n,\mathbf{k}%
\,\nu \,\mu \rangle _{_{A}}.
\end{eqnarray}
Here $t$ is the NN t-matrix; $\nu $ and $\mu $ denote the initial spin and
isospin projections of the projectile and 
\[
\overline{\psi}_{\frac{1}{2}m,\frac{1}{2}n}(\mathbf{p})=(-)^{\frac{1}{2}-m+%
\frac{1}{2}-n}\psi _{\frac{1}{2}-m,\frac{1}{2}-n}(\mathbf{p}) 
\]
where $\psi _{\frac{1}{2}-m,\frac{1}{2}-n}(\mathbf{p})$ annihilates a
nucleon with momentum $\mathbf{p}$ and spin and isospin projections $-m$ and 
$-n$ respectively. The choice of the pair $(\mu ,\mu ^{\prime })$ is
determined by the reaction being considered.

Our study focuses on quasielastic scatering to the isobaric analog state.
Considering explicitly the isospin degrees of freedom for the scattering
waves in the form of outgoing proton and neutron wavefunctions, we obtain a
non-local version of the coupled-channel Lane equations for incoming
protons, 
\begin{equation}
\left( 
\begin{array}{c}
{\Psi _{p}} \\ 
\Psi _{n}%
\end{array}%
\right) =\left( 
\begin{array}{c}
{\phi _{p}} \\ 
0%
\end{array}%
\right) +\left[ 
\begin{array}{ll}
{G_{p}U_{pp}^{(s)}} & G_{p}U_{px} \\ 
G_{n}U_{nx} & G_{n}U_{nn}%
\end{array}%
\right] \left( 
\begin{array}{c}
{\Psi _{p}} \\ 
\Psi _{n}%
\end{array}%
\right) \;.  \nonumber
\end{equation}%
Here we denote by $G_{p}$ ($G_{n}$) the Green functions for charged
(uncharged) outgoing particles. Furthermore, $U_{pp}^{s}$ represents a
short-range interaction where the point-Coulomb interaction has been
subtracted from the hadronic-plus-Coulomb contribution, i.e. $%
U_{pp}^{(s)}=(U_{H}+U_{Ch})-U_{Pt}$. This coupled-channel integral equation
is solved using standard numerical methods. The primary input in these
equations is the optical potential which we obtain following a current
version of the full-folding optical model approach to nucleon scattering 
\cite{Are02,Amo00}. Here we account thoroughly for the Fermi motion of the
target nucleons. The effective interaction in the form of isospin-symmetric
nuclear matter $g$ matrix, is treated fully off-shell. An essential
ingredient in these parameter-free constructions is the nuclear density of
the target. However, intermediate-energy applications for $^{48}$Ca(p,n) and 
$^{90}$Zr(p,n) substantially underestimate the Fermi cross section, with
only marginal differences in the scattering observables when considering
conventional alternative representations of the nuclear densities.

\section{APPLICATIONS AND CONCLUSIONS}
In order to explore this difficulty we have devised model-independent
densities which, once folded with the \emph{in-medium} effective
interactions, generate an optical potential. These model-independent
representations of the density have the general form 
\begin{equation}
\rho _{p,n}(r)=\rho _{p,n}^{\circ }(r)(1+\textstyle{\frac{1}{2}}{\xi
_{p,n}(r)})^{2}\;,  \nonumber
\end{equation}%
where $\rho _{p,n}^{\circ }(r)$ is a reference density taken from a
reasonable model (3pF, 3pG, etc.) and $\xi _{p,n}(r)$ is a dressing function
constructed from an $N$-knot spline. The value of $\xi $ at each of the
selected knots becomes a searchable parameter. We have allowed variations of
both proton and neutron densities in order to fit in the best possible way
the measured cross sections from quasielastic charge-exchange experiments.
For physical consistency we have constrained the full proton densities to
the measured charge r.m.s radius from electron scattering experiments (and
the volume integrals of $\rho _{n}$ and $\rho _{p}$ to give N and Z
respectively). \\
\begin{figure}[htb]
\vspace{-.8cm}
\begin{minipage}[t]{75mm}
\includegraphics[width=75mm]{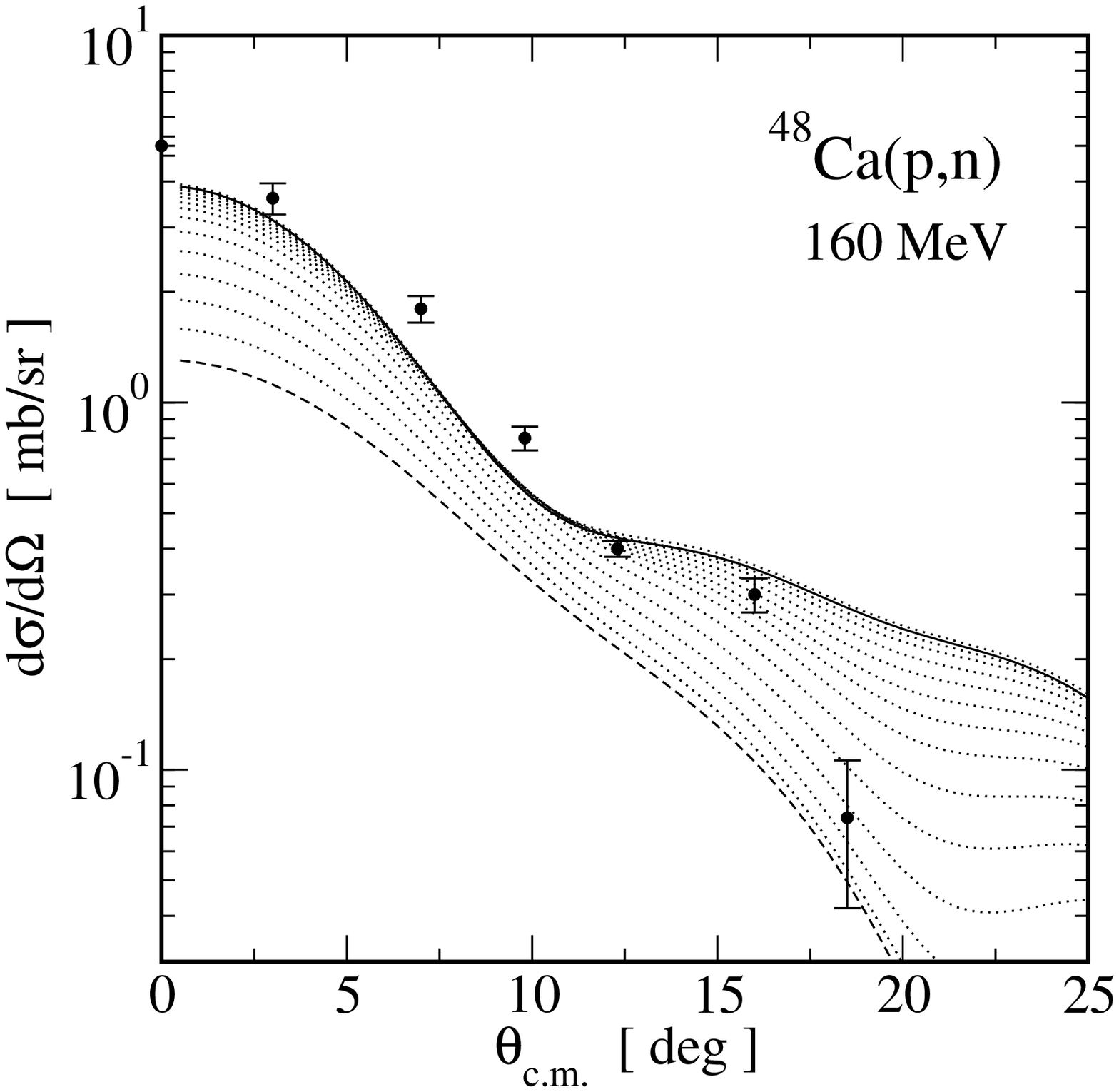}
\vspace{-.5cm}
\caption{
Differential cross-section for $^{48}$Ca(p,n) quasielastic
scattering at 160 MeV.
The data are taken from Ref. \cite{Rap}.
The solid curve corresponds to the last iteration. 
}
\label{fig1}
\end{minipage}
\hspace{\fill}
\begin{minipage}[t]{75mm}
\includegraphics[width=75mm]{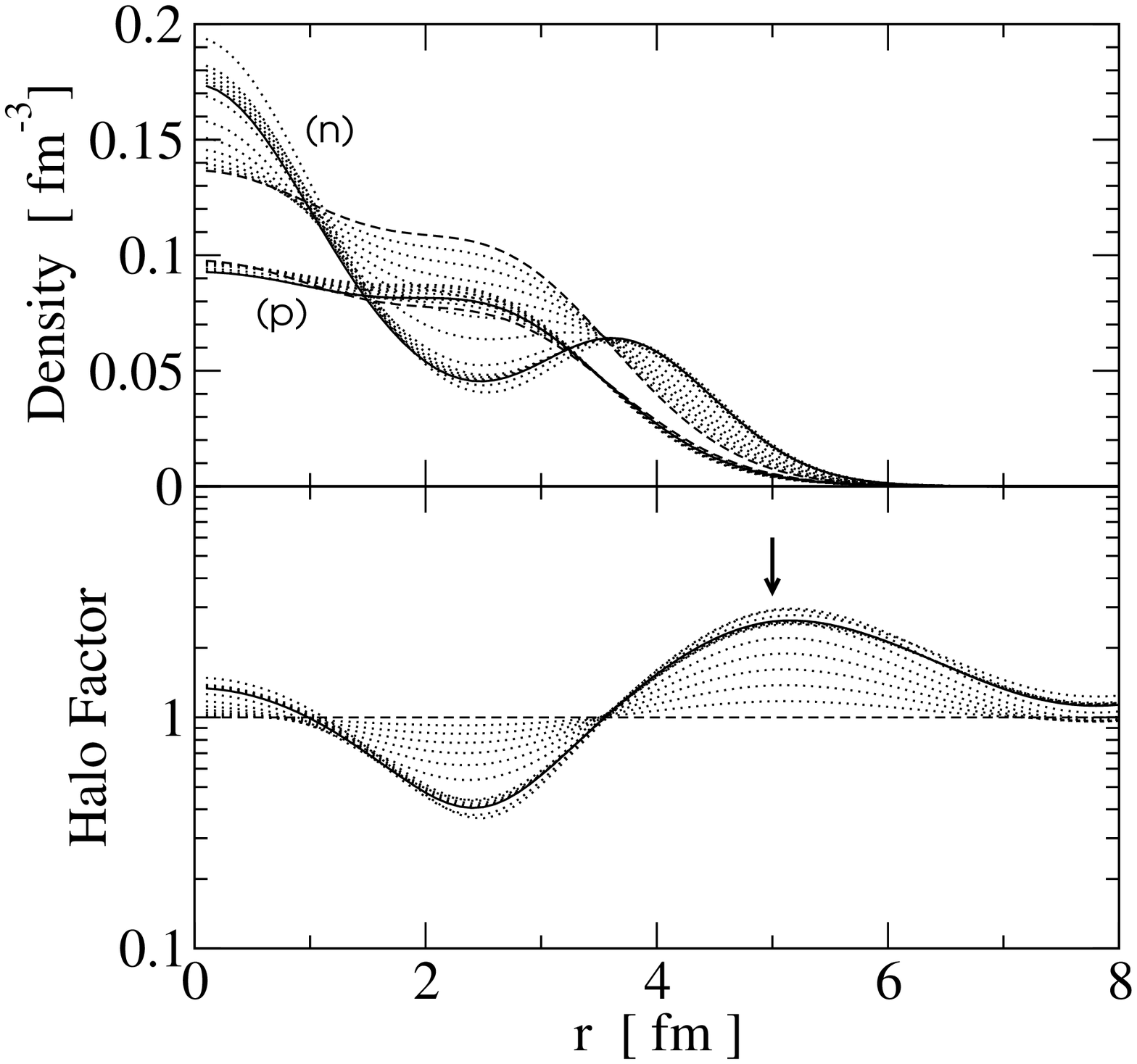}
\vspace{-.5cm}
\caption{
Evolution of the proton (p) and neutron (n) point
densities (upper frame), and halo factor (lower frame) in search of the best
fit of the measured forward-angle differential cross section. Results from
the last iteration are shown with solid curves.
}
\label{fig2}
\end{minipage}
\end{figure}

In Fig. 1 we show results for the iterative search in the
case of $^{48}$Ca(p,n) at 160 MeV. Each curve represents the calculated
differential cross-section for a given density. 
In Fig. 2 we
represent the corresponding proton (p) and neutron (n) densities, and
respective halo factor defined as $f(r)=Z\rho _{n}(r)/N\rho _{p}(r)$. It
is interesting to observe the manifestation of a halo structure at a
distance near 4 fm. Indeed, a close examination of the obtained results
indicates a peripheral halo factor of 2.3, in good agreement with values
reported in Ref. \cite{Sch03}. 
Another quantity of interest in the
characterization of neutron halos is the nuclear skin, defined as the
difference between the neutron and proton r.m.s. radii. We obtain 0.52 fm
for this quantity, which is much larger than conventional reported values.

Here we have focused on variations in the proton and neutron densities; a
closer examination of these results may require additional constraints from
the elastic channel, a closer scrutiny of the isovector strength and/or the
inclusion of higher order effects such as asymmetric nuclear matter
effective interactions.

\acknowledgements{
Partial financial support provided by FONDECYT 
under grant No 1040938.
}

\end{document}